\documentstyle[aps,12pt]{revtex}
\begin{document}
{\normalsize
\title{Critical temperature and density of spin-flips in the
anisotropic random field Ising model\\[0.5cm]}
\author{Marc~Thilo~Figge, Maxim~V.~Mostovoy\cite{Perm}, and
Jasper~Knoester
}
\address{\normalsize Institute for Theoretical Physics and 
Materials Science Center\\
University of Groningen, Nijenborgh 4, 9747 AG Groningen,
The Netherlands}
\date{\today}
\maketitle
\baselineskip 28pt
%
\begin{abstract}

\normalsize
\baselineskip 28pt
We present analytical results for the strongly anisotropic random
field Ising model, consisting of weakly interacting spin chains.
We combine the mean-field treatment of interchain interactions
with an analytical calculation of the average chain free energy
(``chain mean-field'' approach).  The free energy is found using
a mapping on a Brownian motion model.  We calculate the order
parameter and give expressions for the critical random magnetic
field strength below which the ground state exhibits long range
order and for the critical temperature as a function of the
random magnetic field strength.  In the limit of vanishing
interchain interactions, we obtain corrections to the
zero-temperature estimate by Imry and Ma [Phys.~Rev.~Lett.~{\bf
35}, 1399 (1975)] of the ground state density of domain walls
(spin-flips) in the one-dimensional random field Ising model.
One of the problems to which our model has direct relevance is
the lattice dimerization in disordered quasi-one-dimensional
Peierls materials, such as the conjugated polymer {\em
trans}-polyacetylene.
\end{abstract}
PACS numbers: 75.10.-b, 05.50.+q, 64.60.Cn
%

\section{Introduction}\label{intro}

%
Owing to the wide range of physical systems that can be described
by Ising-like models, it is believed that the random field Ising
model (RFIM), {\em i.e.}, the Ising model with a spatially random
magnetic field, captures the essential physics of many disordered
systems.  Therefore, over the past two decades this model has
been extensively studied, both theoretically \cite{Nattermann97}
and experimentally \cite{Belanger97}.  The most studied
realization of the RFIM is the dilute antiferromagnet in a
uniform field.  This realization was first found by Fishman and
Aharony \cite{Fishman79} in 1979 and is still under current
investigation \cite{Hill97,Feng97}.  Other equally interesting as
diverse problems to which the RFIM has been found to be
applicable are, {\em e.g.}, the Anderson-Mott transition of
disordered interacting electrons \cite{Kirkpatrick96} and aspects
of protein folding \cite{Gutin96}.

Much of the theoretical interest in the RFIM has originated from
the early work by Imry and Ma \cite{Imry75}.  In 1975, these
authors noticed that in the presence of a random magnetic field
the lowest energy state of the Ising model may have no long range
order (LRO).  They considered the energetics of creating a
``wrong'' domain ({\em i.e.}, a domain in which all spins have
reversed signs) of linear size $R$ in the ordered phase.  The
typical domain size $R$ caused by the disorder fluctuations is
determined by the balance between the energy cost of domain walls
and the possible energy gain due to the decrease of the
interaction energy between the reversed spins and the random 
field. In the one-dimensional case, this balance reads:
\begin{equation}
2 I\;\sim\;2 \sqrt{\langle h^2 \rangle R} ,
\label{energyra}
\end{equation}
where $I$ is the energy of one domain wall and
$\langle h^2 \rangle$ is the strength of the random magnetic
field.
This
zero temperature estimate implies that in the one-dimensional
random field Ising model LRO does not exist, as in an infinite
system it is always possible to choose $R$ large enough to make
the creation of a ``wrong'' domain favorable.  From
Eq.(\ref{energyra}), the zero temperature density of spin-flips
is estimated to be \cite{Imry75,Imry84}:
\begin{equation}
n_{s}\;=\;\frac{1}{R}\;\sim\;\frac{\langle
h^{2}\rangle}{I^{2}} .
\label{imryma}
\end{equation}
The lower critical dimension above which LRO in the RFIM is
possible is $2$ \cite{Imry75,Imbrie84}.

In most previous studies of the RFIM, the interactions between
the Ising spins were assumed to be isotropic, {\em i.e.},
independent of the direction.  These studies focused on the
universal critical properties of the model, such as the critical
exponents and the lower critical dimension, which do not depend
on the degree of anisotropy of the interactions.  On the other
hand, the anisotropic RFIM has several nonuniversal properties,
such as the critical temperature, that are very interesting in
relation to disordered quasi-one-dimensional systems.

For example, we recently showed \cite{Mostovoy97} that disordered
Peierls chains with a doubly degenerate ground state can under
certain conditions be described by the one-dimensional RFIM.  In
this mapping, the Ising variables describe the local lattice
dimerization, while the random field corresponds to disorder in
the electron hopping amplitudes.  The spin-flips induced by the
random field then correspond to disorder-induced neutral solitons
in Peierls chains.  The density of spin-flips is related to the
magnetic properties of the disordered Peierls systems, since the
neutral solitons have spin $1/2$ \cite{SSHK}.  The conjugated
polymer {\em trans}-polyacetylene is a well-known example of a
Peierls system for which this mapping may be used.  Taking into
account the weak interchain interactions in quasi-one-dimensional
crystalline {\it trans}-polyacetylene, the thermodynamics of
solitons in this material may be described by the anisotropic
RFIM \cite{Figge97}.  In this context, the interesting properties
of the anisotropic RFIM are the temperature of the
three-dimensional Peierls transition, the order parameter, and
the density of spin-flips as a function of the random field
strength and (or) temperature.  These properties strongly depend
on the degree of anisotropy of the interactions.

In the present paper we study the $d$-dimensional anisotropic
RFIM ($d > 2$), which consists of regularly arranged chains of
spins in a random magnetic field (Sec.~\ref{mod}).  The
interaction between spins in neighboring chains is supposed to be
much smaller than the interaction within chains and will be
treated in the mean-field approximation; the intrachain
interactions will be treated exactly.  Using the transfer matrix
formalism and a mapping on a Brownian motion model, we find an
analytical expression for the average free energy of the
continuum version of our model (Sec.~\ref{mapping}).  This key 
result is used in Sec.~\ref{anal} to obtain the order parameter, 
the critical temperature, and the density of spin-flips.  
In Sec.~\ref{aIM}, we present analytical results for these 
quantities in the limit where thermal creation of spin-flips 
dominates their creation by the random magnetic field; in 
Sec.~\ref{aRFIM} we consider the opposite limit.  
We find the critical random field strength below
which the ground state exhibits LRO; well below the critical
strength, the density of spin-flips is found to be exponentially
suppressed.

In the absence of the interchain interactions, the anisotropic
model reduces to the one-dimensional RFIM.  In this limiting
case, our solution is exact.  It shows that the estimate for the
density of spin-flips Eq.(\ref{imryma}) in a single chain is
correct to lowest order in the random magnetic field strength and
gives higher-order corrections to this expression.  Previous
attempts to obtain an exact solution for the one-dimensional case
\cite{Paladin92} did not yield closed expressions for the free
energy and the spin-flip density.

We conclude this paper by giving the physical interpretation of
our results in Sec.~\ref{discussion} and summarizing in
Sec.~\ref{concl}.

\section{Model}\label{mod}


We consider the anisotropic RFIM on a $d$-dimensional lattice
with a total number of $N^d$ sites.  The Ising spins occupying
the lattice, interact with their nearest neighbors and with a
magnetic field that has a random value at each lattice site.  The
anisotropy results from the fact that the exchange interaction,
$I$, along one of the lattice directions is assumed to be much
stronger than the interaction, $I_{\perp}$, perpendicular to this
direction: $0\;<\;I_{\perp}\;\ll\;I$ (for definiteness, the
interactions are chosen positive).  The lattice is thus divided
into $N^{d-1}$ chains (labeled $\alpha$) of $N$ spins (labeled
$i$).  The energy of our model is given by:
\begin{equation}
{\cal E}[h]\;=\;
I \!\sum_{\langle i, i^{\prime}\rangle, \alpha }\,
\:\frac{1-\sigma_{i \alpha}\,\sigma_{i^{\prime} \alpha}}{2}
+ I_{\perp}\!\sum_{i, \langle \alpha, \alpha^{\prime}\rangle }\,
\:\frac{1-\sigma_{i \alpha}\,\sigma_{i \alpha^{\prime}}}{2}
 - \sum_{i, \alpha}\,h_{i \alpha}\,\sigma_{i \alpha} .
\label{action}
\end{equation}
Here, $\sigma_{i \alpha}=\pm 1$ is the classical spin variable
describing the two possible spin directions, up and down, at site
$i$ of chain $\alpha$.  The summations over $\langle i,
i^{\prime}\rangle$ and $\langle \alpha, \alpha^{\prime}\rangle$
are restricted to nearest neighbor pairs.  The random magnetic
field $h_{i \alpha}$ is assumed to be uncorrelated for different
lattice sites, with vanishing mean and Gaussian correlator of
strength $\epsilon$:
\begin{equation}
\langle h_{i \alpha}\rangle\,=\,0\;\;\;\;{\rm and}\;\;\;\;
\langle h_{i \alpha}\,h_{j \beta}\rangle\;=
\;\epsilon\:\delta_{i,j}\,\delta_{\alpha, \beta} .
\label{gausscorr}
\end{equation}
Here, $\langle \ldots \rangle$ denotes the average over the
random field realizations.  Finally, we impose open boundary
conditions on the model.  This choice does not affect our
results, as these will be derived for a large system,
$N\rightarrow\infty$.

For vanishing magnetic field strength $\epsilon$, the ground
state of the spin system Eq.(\ref{action}) is ferromagnetically
ordered.  This is not necessarily the case when a finite random
field is present: For a number of spins it may then be favorable
to flip and align with the local direction of $h_{i \alpha}$ to
minimize the total energy.

To calculate the partition function of the model we treat the
interchain interactions in the mean-field approximation. Then
the energy Eq.(\ref{action}) becomes a sum of single-chain
energies and can be written as
\begin{equation}
{\cal E}_{MF}[h]\;=\;N^{d-1} \sum_i \left[
\frac{I}{2} \left(1 - \sigma_{i} \sigma_{i+1}\right)
- (B + h_i) \sigma_i + \frac{B^2}{{\tilde I}_{\perp}} +
\frac{{\tilde I}_{\perp}}{4} \right] ,
\label{EMF}
\end{equation}
where $B$ is the homogeneous mean field, ${\tilde I}_{\perp}$
equals $I_{\perp}$ times the number of nearest-neighbour chains
of one particular chain, and we took into account that the
contributions of all $N^{d-1}$ chains are on average equal.  The
mean field $B$ satisfies the self-consistency equation:
\begin{equation}
B\;=\;\frac{1}{2}\,{\tilde I}_{\perp}\,
\langle\langle\sigma\rangle\rangle ,
\label{Bmf}
\end{equation}
where  $\langle\langle \ldots \rangle\rangle$ denotes both the
thermal average and the average over random field realizations.
The order parameter $\langle\langle\sigma\rangle\rangle$, which
is the normalized difference between the average number of sites
with spin up ($N_{\uparrow}$) and spin down ($N_{\downarrow}$),
\begin{equation}
\langle\langle\sigma\rangle\rangle\;=\;
\frac{N_{\uparrow}\,-\,N_{\downarrow}}
{N_{\uparrow}\,+\,N_{\downarrow}} \;,
\end{equation}
can be found from
\begin{equation}
\langle\langle\sigma\rangle\rangle\;=\;-
\frac{1}{N} \frac{\partial}{\partial B}
\left\langle F \right\rangle .
\label{order}
\end{equation}
Here, $\left\langle F \right\rangle$ denotes the average over the
random field realization of the free energy of a single chain in
the mean field $B$. For a particular random field realization:
\begin{equation}
F[h]\;=\;
- \frac{1}{\beta} \ln \left(
\sum_{\{\sigma\}}\,e^{- \beta E[\sigma,h]} \right) ,
\label{partfunc}
\end{equation}
where the chain energy $E[\sigma,h]$ is given by the relevant
part of Eq.(\ref{EMF}),
\begin{equation}
E[\sigma,h]\;=\;\sum_i \left[
\frac{I}{2} \left(1 - \sigma_{i} \sigma_{i+1}\right)
- (B + h_i) \sigma_i \right] .
\label{mfaction}
\end{equation}
Using Eq.(\ref{order}), the explicit form of the self-consistency
condition (\ref{Bmf}) becomes:
\begin{equation}
B\;=\;-
\frac{{\tilde I}_{\perp}}{2N} \frac{\partial}{\partial B}
\left\langle F \right\rangle .
\label{newsc}
\end{equation}
In view of the relative weakness of the interchain interactions
($I_{\perp} \ll I$), this so-called chain mean-field
approximation\cite{IPS} is expected to be accurate.

The disorder-averaged free energy, $\left\langle F
\right\rangle$, which we calculate analytically in the next
section, can also be used to find other relevant thermodynamic
properties. In particular, the average density of spin-flips
inside a chain as a function of temperature and random field
strength is obtained through:
\begin{equation}
n_s\;=\;
\frac{1}{N}\frac{\partial}{\partial I}
\left\langle F \right\rangle ,
\label{densform}
\end{equation}
since $I$ is the spin-flip creation energy. In the context of the
degenerate ground state conjugated polymers, mentioned in the
Introduction, $n_s$ directly gives the density of solitons within
chains.


\section{The average free energy}\label{mapping}


To derive the continuum version of the discrete model
Eq.(\ref{mfaction}), we follow the standard path
\cite{Fradkin78,Kogut79} by first rewriting the chain partition
function in the transfer matrix formalism.  The partition
function is thus considered as the expectation value of an
ordered product of operators $\hat{T}_{i}$ along the chain axis:
\begin{equation}
Z[h]\;=\;\sum_{\{\sigma\}}\,
\exp\left\{\,- \beta E[\sigma,h]\,\right\}\;=\;
\sum_{\sigma,\sigma^{\prime}=\pm 1}
\,\langle\sigma^{\prime}\,|\,\prod^{N}_{i=1}\,
\hat{T}_{i}\,|\,\sigma\,\rangle .
\label{tform}
\end{equation}
Here, the final summation over $\sigma,\sigma^{\prime}$ accounts
for all possible boundary conditions for the spins on the chain's
ends.  On the basis spanned by the two vectors $|\,+\,\rangle$
and $|\,-\,\rangle$ (corresponding to $\sigma=+1$ and
$\sigma=-1$, respectively), the transfer matrix $\hat{T}_{i}$
reads:
\begin{eqnarray}
&&\nonumber\\
\hat{T}_{i}\;&=&\;
\left(\begin{array}{cc}
e^{+\beta (h_{i}+B)} &
e^{-\beta I}\,e^{+\beta (h_{i}+B)}\\ \\
e^{-\beta I}\,e^{-\beta (h_{i}+B)} &
e^{-\beta (h_{i}+B)}
\end{array} \right)\;\;.\label{tmatrix}\\
&&\nonumber
\end{eqnarray}
$\hat{T}_{i}$ depends on the site-index $i$ through the random magnetic
field.

A relation between the transfer matrix Eq.(\ref{tmatrix}) and its
(euclidean) quantum Hamiltonian is established by interpreting
the axis of the chain as the (imaginary) time axis of quantum
mechanics.  This means that neighboring sites labeled by $i$
are considered as subsequent times $t$ in the continuum model.
In this sense $\hat{T}_{i}$ carries information about the time
evolution of the system and can in fact be identified with the
time evolution operator $\hat{T}$ of a corresponding Hamiltonian
$\hat{H}$:
\begin{equation}
\hat{T}\;=\;\exp\left\{-\hat{H} \right\} .
\label{tordop}
\end{equation}
In the continuum approximation the deviation of $\hat{T}$ from
the identity matrix is assumed to be small, so that the last
expression can approximately be written as
\begin{equation}
\hat{T}\;=\;1\,-\,\hat{H} ,
\label{taucont}
\end{equation}
where $\hat{H}$, obtained from Eq.(\ref{tmatrix}), has the form:
\begin{equation}
\hat{H}(t)\;=\;\beta\,(h(t)\,+\,B)\,\hat{\sigma}_{3}\;+\;
\exp\{-\beta I\}\,\hat{\sigma}_{1} .
\label{euquaham}
\end{equation}
Here, $\hat{\sigma}_{1}$ and $\hat{\sigma}_{3}$ are the
Pauli matrices and the random magnetic field is now Gaussian in
time:
\begin{equation}
\langle h(t) \rangle\,=\,0\;\;\;\;{\rm and}\;\;\;\;
\langle h(t)\,h(t^{\prime})
\rangle\;=\;\epsilon\,\delta(t-t^{\prime}) .
\label{timegausscorr}
\end{equation}
In the last expression, we have chosen the lattice constant as
the unit of length (time).
The Hamiltonian Eq.(\ref{euquaham}) describes the relaxation of a
spin $1/2$ in a magnetic field which has a constant $x$-component
and a random and time-dependent $z$-component.

It can easily be seen from Eq.(\ref{tmatrix}) that the validity
of the expansion Eq.(\ref{taucont}), {\em i.e.}, the validity of
the continuum description of the discrete Ising model, requires
the two conditions
\begin{equation}
\exp\{-\beta I\}\;\ll\;1
\label{cond1}
\end{equation}
and
\begin{equation}
\beta\,(|h(t)|\,+\,|B|)\;\ll\;1
\label{cond2}
\end{equation}
to be fulfilled simultaneously.  While the first condition
results in an upper limit for the temperature, $T\ll
I$, the latter implies a lower temperature limit:
$T \gg O\Big({\rm max}\Big\{\sqrt{\epsilon},|B|\Big\}\Big)$.

The continuum limit allows us to write the partition function
Eq.(\ref{tform}) as a sum over matrix elements for the
time-ordered propagation of a spin-1/2 particle in imaginary
time:
\begin{equation}
Z[h]\;=\;
\sum_{\sigma, \sigma^{\prime}=\pm 1}\,
\langle\,\sigma^{\prime}\,|\,
\hat{{\cal T}}\,\exp\left\{\,\int_{t_{i}}^{t_{f}}
\hat{H}(t)\,dt\,\right\}\,|\,\sigma\,\rangle .
\label{pfmap2}
\end{equation}
Here, $\hat{{\cal T}}$ is the time-ordering operator
and the integration over the dimensionless
time-variable $t$ has to be performed from the initial time
$t_{i}=0$ to the final time $t_{f}=N$ corresponding to the
chain length.

It is now convenient to rotate the spinor basis over an angle
$\pi/2$ around the $\sigma_{2}$-axis, which yields the
transformed Hamiltonian:
\begin{eqnarray}
\hat{H}^{\prime}(t)\;&=&\;
\exp\left\{i\frac{\pi}{4}\hat{\sigma}_{2}\right\}\,\hat{H}(t)\,
\exp\left\{-i\frac{\pi}{4}\hat{\sigma}_{2}\right\}\nonumber\\
\;&=&\;-\beta\,(h(t)\,+\,B)\,\hat{\sigma}_{1}\;+\;
\exp\{-\beta I\}\,\hat{\sigma}_{3} ,
\label{rotham}
\end{eqnarray}
and
\begin{equation}
Z[h]\;=\;
2 \,\langle\,+\,| \,\hat{{\cal T}}\,\exp\left\{\,
\int_{0}^{N}\hat{H}^{\prime}(t)dt\,\right\} |\,+\,\rangle .
\label{rotpf}
\end{equation}
To make this expression more compact, we introduce the wave
function of the spin,
\begin{equation}
\Psi(t)\;=\;
\psi_{\uparrow}(t)\,|\,+\,\rangle\;+
\;\psi_{\downarrow}(t)\,|\,-\,\rangle\;,
\label{spinwave}
\end{equation}
which obeys the time-dependent Schr\"{o}dinger-like equation,
\begin{equation}
\frac{d}{dt}\,\Psi(t)\;=\;\hat{H}^{\prime}(t)\,\Psi(t)\;.
\label{schrodinger}
\end{equation}
Together with the inital condition $\Psi(t=0)=|\,+\,\rangle$, the
partition function Eq.(\ref{rotpf}) now simply reads:
\begin{equation}
Z [h]\;=\;2
\,\psi_{\uparrow}(N) .
\label{pfcompact}
\end{equation}

Our problem of calculating the partition function has, of course,
not been solved in going from Eq.(\ref{tform}) to
Eq.(\ref{pfcompact}).  However, it is possible to find a
substitution for the spin wave function Eq.(\ref{spinwave}) that
relates the calculation of $Z[h]$ to an exactly
solvable Brownian motion model.  This method was already used by
us in a previous paper \cite{Mostovoy97}, where we considered the
Hamiltonian Eq.(\ref{euquaham}) without interchain interactions
($B=0$), in the context of disorder-induced solitons in a single
polymer chain.  To keep the present paper reasonably
self-contained, we will briefly repeat the essential steps of
this method, allowing now also for finite $B$ values.  Even though
from a technical point of view, having $B \ne 0$ is not a major
complication, the physical consequences are important, as it gives
rise to the existence of LRO at finite temperature and (or) finite
random magnetic field.

We start by introducing the two time-dependent functions $u(t)$
and $v(t)$, such that the spin wave function Eq.(\ref{spinwave})
reads:
\begin{equation}
\Psi(t)\;=\;
u(t)\,\left(\begin{array}{c}
\cosh(\frac{v(t)}{2})\\ \,\\
\sinh(\frac{v(t)}{2})
\end{array}\right) .
\label{substi}
\end{equation}
The initial condition, $\Psi(t=0)=|\,+\,\rangle$, requires that
$u(t=0)=1$ and $v(t=0)=0$, and from Eq.(\ref{schrodinger}) we
obtain two coupled differential equations for $u(t)$ and $v(t)$,
\begin{equation}
\frac{du(t)}{dt}\;=\;u(t)\,e^{-\beta I}\,\cosh(v(t)) ,
\label{sub1}
\end{equation}
and
\begin{equation}
\frac{dv(t)}{dt}\;=\;-\,2\,\beta\,B\,-\,
2\,e^{-\beta I}\,\sinh(v(t))\,-\,
2\,\beta\,h(t) .
\label{sub2}
\end{equation}
While Eq.(\ref{sub1}) can be easily integrated, yielding
\begin{equation}
u(t)\;=\;u(0)\,\exp\left\{\,e^{-\beta I}
\int_{0}^{t}\cosh(v(t^{\prime}))\,dt^{\prime}\,\right\} ,
\label{uint}
\end{equation}
Eq.(\ref{sub2}) can be interpreted as the Langevin equation for
the velocity $v$ of a particle undergoing a one-dimensional
Brownian motion.  The last term in Eq.(\ref{sub2}) is the random
force acting on the particle, while the first and the second terms
describe, respectively, a constant (bias) force and a friction.

In the absence of the random force (no random magnetic field),
the velocity $v$ approaches a time-independent value for $t
\rightarrow \infty$:
\begin{equation}
v_{\infty}\;=\;- {\rm arcsinh}\Big(\,\beta\,B\,e^{\,\beta
I}\,\Big) .
\label{simpvelo}
\end{equation}
In the presence of the random force, $v(t)$ becomes a stochastic
variable with a time-dependent distribution, $P(v,t)$, which may
be obtained from the corresponding Fokker-Planck equation
\cite{Kubo95}.  In the limit of long times ($N\rightarrow
\infty$), the time-independent equilibrium distribution,
$P_{eq}(v)$, is reached, obeying:
\begin{equation}
-\frac{\partial}{\partial v}\Big(\,
\alpha_{1}(v)\,P_{eq}(v)\,\Big)\;+\;
\frac{\alpha_{2}}{2}\,
\frac{\partial^{2} P_{eq}(v)}{\partial v^{2}}\;=\;
\frac{\partial P_{eq}(v)}{\partial t}\;=\;0 .
\label{fpequ}
\end{equation}
Here, the first and second moments are given by, respectively,
\begin{equation}
\alpha_{1}(v)\;=\;
-2 \beta B\;-\;2\,e^{-\beta I}\,\sinh(v(t)) ,
\label{deltaone}
\end{equation}
and
\begin{equation}
\alpha_{2}\;=\;4 \beta^2
\int_{0}^{N}\left\langle h(t)
h(t^{\prime}) \right\rangle dt^{\prime}
\;=\;4\,\beta^{2}\,\epsilon .
\label{deltatwo}
\end{equation}
The solution of the Fokker-Planck equation
(\ref{fpequ}) is easily found to be:
\begin{equation}
P_{eq}(v)\;=
\;{\cal N}\,
\exp\bigg\{- \nu\,v - z \cosh(v)\bigg\} ,
\label{fpsol}
\end{equation}
where we introduced
\begin{equation}
\nu\;=\;\frac{B}{\beta \epsilon} ,
\label{besselorder}
\end{equation}
\begin{equation}
z\;=\;\frac{e^{-\beta I}}{\beta^{2} \epsilon} ,
\label{argument}
\end{equation}
and the normalization coefficient ${\cal N}$ determined by:
\begin{equation}
\int_{-\infty}^{+\infty}dv\,P_{eq}(v)\;=\;1 .
\label{normcoeff}
\end{equation}
The equilibrium distribution Eq.(\ref{fpsol}) is centered around
$v_{\infty}$ (Eq.(\ref{simpvelo})), and in the limit of vanishing
fluctuations ($\epsilon\rightarrow 0$), it approaches a
$\delta$-distribution [$P_{eq}(v)
\rightarrow\;\delta(v\,-\,v_{\infty})$].

Using our result for $P_{eq}(v)$, the average free energy
of the RFIM can easily be obtained. We first note that
Eq.(\ref{uint}) enables us to write the partition function
Eq.(\ref{pfcompact}) in terms of $v(t)$ alone:
\begin{eqnarray}
Z[h]\;&=&\;2
\,\psi_{\uparrow}(N)
\;=\;2
\,u(N)\,\cosh\left(\frac{v(N)}{2}\right)\;=\;\nonumber\\
\;&=&\;2
\,\exp\left\{\,e^{-\beta I}
\int_{0}^{N}\cosh(v(t))\,dt\,\right\}\,
\cosh\left(\frac{v(N)}{2}\right) .
\label{langpf}
\end{eqnarray}
Omitting in the limit of long chains all terms that do not
grow proportional to $N$,
we can now replace the average of $\ln Z[h]$ over the
random field realizations by the average over $P_{eq}(v)$. 
This leads to
\begin{equation}
\frac{1}{N}\,\left\langle\,F\,\right\rangle\;=\;
-\,\frac{e^{-\beta I}}{\beta}\,
\int_{-\infty}^{+\infty}\!\!P_{eq}(v)\cosh(v)\,dv\;=\;
B\,\frac{z}{\nu}\,
\frac{K^{\prime}_{\nu}(z)}{K_{\nu}(z)} \;,
\label{vaverage}
\end{equation}
where $K_{\nu}(z)$ denotes the modified Bessel function of order
$\nu$ and $K_{\nu}^{\prime}(z)$ its derivative with respect to
$z$ \cite{Abramowitz}.

The analytical solution Eq.(\ref{vaverage}) of the average
free energy is the key result of this paper, from which all
further results follow. In Fig.~1, this solution as a function
of temperature is compared to results of a numerical simulation
of the discrete RFIM at $\epsilon = 0.01 I^2$ for three values of
the mean field $B$. Apart from the presence of the mean field,
the simulation method is identical to the one described in
Ref.~\cite{Mostovoy97}. Figure~1 shows that in the temperature
region $T \ll I$, the analytical solution is in excellent
agreement with the simulation, except for very low temperatures,
where the condition Eq.(\ref{cond2}) is not met. It is observed
that the analytical solution for the free energy reaches a
maximum at $T = T_0(\epsilon) \neq 0$ (indicated by the thin
line). The meaning of $T_0(\epsilon)$ will be discussed in
Sec.~\ref{aRFIM}.


\section{Analysis of results}
\label{anal}

Using Eq.(\ref{vaverage}), it is straightforward to solve
numerically the self-consistency equation (\ref{newsc}), from
which the order parameter $\langle\langle \sigma \rangle\rangle$
is found through Eq.(\ref{Bmf}).  Figure~2 gives the thus
obtained order parameter as a function of temperature and
random-field strength for ${\tilde I}_{\perp} / I = 0.02$.  A
low-temperature cut-off $T_0(\epsilon)$ was used, as will be
explained in detail in Sec.~\ref{aRFIM}.  The thick curve in the
$\langle \langle \sigma \rangle\rangle = 0$ plane separates the
region with LRO (inside this curve) from the one without LRO.
From the solution for $\langle \langle \sigma \rangle\rangle$,
one obtains the critical temperature and the density of
spin-flips through further numerical analysis.  For certain
parameter values $\nu$ and $z$, however, our result
Eq.(\ref{vaverage}) also allows for analytical solutions.  Here,
two limiting cases should be distinguished.  In Sec.~\ref{aIM},
we will consider the case of very weak disorder, where thermally
induced spin-flips dominate the ones induced by the random
magnetic field.  In Sec.~\ref{aRFIM}, we consider the opposite
limit.


\subsection{Dominant thermal spin-flips}
\label{aIM}

%

For a weak random magnetic field ($\epsilon\rightarrow 0$) the
modified Bessel function in Eq.(\ref{vaverage}) has to be
evaluated at large order and argument $(\nu, z \gg 1)$. Using
standard expansions \cite{Abramowitz}, we obtain to first order
in $\epsilon$:
\begin{equation}
\frac{1}{N} \left\langle  F \right\rangle =
- \frac{1}{\beta}
\sqrt{\left(\beta B\right)^2
+ e^{-2 \beta I}} -
\frac{\beta \epsilon}{2}
\frac{e^{- 2 \beta I}}
{\left( \beta B \right)^2 + e^{- 2 \beta I}} \;.
\label{simpfree}
\end{equation}
The first term in the right hand side of Eq.(\ref{simpfree}) is
the continuum approximation for the free energy of the discrete
model Eq.(\ref{mfaction}) at $\epsilon = 0$. Under the conditions
specified by Eqs.(\ref{cond1}) and (\ref{cond2}) this term
agrees with the exact expression,
\begin{equation}
\frac{1}{N} F =
- \frac{1}{\beta} \ln \left( \cosh \left( \beta B \right) +
\sqrt{\left(\sinh \left(\beta B\right) \right)^2
+ e^{-2 \beta I}}\right) ,
\end{equation}
which can be obtained by diagonalization of the
transfer-matrix Eq.(\ref{tmatrix}) at $h_i = 0$.

Using Eq.(\ref{simpfree}), the self-consistency equation
(\ref{newsc}) for the mean field $B$ can be written in the form:
\begin{equation}
\frac{2T}{{\tilde I}_{\perp}} =
y - \frac{\epsilon}{T^2} e^{-2 \beta I} y^4 ,
\label{scweak}
\end{equation}
where
\begin{equation}
y = \left[ \left(\beta B \right)^2 + e^{- 2 \beta I}
\right]^{-\frac{1}{2}} .
\end{equation}
At the transition temperature, $T_c(\epsilon)$, the order
parameter vanishes.  Therefore, by putting $B = 0$ in
Eq.(\ref{scweak}), we obtain an equation for $T_c(\epsilon)$:
\begin{equation}
\frac{2 T_c(\epsilon)}{{\tilde I}_{\perp}} =
\exp\left(\frac{I}{T_c(\epsilon)}\right) -
\frac{\epsilon}{T_c^2(\epsilon)}
\exp\left( 2 \frac{I}{T_c(\epsilon)}\right) .
\label{Tc(eps)}
\end{equation}

In particular, at $\epsilon = 0$ we obtain:
\begin{equation}
\frac{{\tilde I}_{\perp}}{2 T_{c}(0)}
\exp \left(\frac{I}{T_c(0)}\right) =
1 \;.
\label{tcsimpl}
\end{equation}
This equation can be compared with the exact result for the
critical temperature of the two-dimensional Ising model
\cite{Onsager,Baxter82} :
\begin{equation}
\sinh \left(\frac{I_{\perp}}{T_c} \right)
\sinh \left(\frac{I}{T_c} \right) = 1 .
\label{2DIsing}
\end{equation}
For $I_{\perp} \ll I$, both Eq.(\ref{tcsimpl}) and
Eq.(\ref{2DIsing}) give approximately:
\begin{equation}
T_c \sim \frac{I}
{\ln\left(\frac{I}{I_{\perp}}\right)} \;,
\end{equation}
which shows that the chain mean-field approximation works well
for the strongly anisotropic Ising model.

From Eq.(\ref{Tc(eps)}) we find that at weak disorder the
transition temperature decreases linearly with $\epsilon$,
\begin{equation}
T_c(\epsilon) = T_c(0) \left(1 - \alpha \epsilon \right) ,
\label{Tclinear}
\end{equation}
where $\alpha$ is:
\begin{equation}
\alpha = \frac{2}{{\tilde I}_{\perp}\left(I + T_c(0)\right)} \;.
\end{equation}

Finally, the average density of spin-flips
[see Eq.(\ref{densform})] for weak disorder is given by
\begin{equation}
n_s = \left\{
\begin{array}{l@{\quad \quad}l}
e^{-\beta I} & {\rm for}\;\; T > T_{c}(\epsilon) \nonumber \\ &
\;\;\;\;\;\;\;\;\;\;\;\;\;\;\;\;\;\;\;\;\;\; . \label{denssimpl}
\\ e^{- 2 \beta I}\left(\frac{2}{{\tilde I}_{\perp}\,\beta}
+ \epsilon \beta^2 y^2\right) & {\rm for}\;\;T < T_{c}(\epsilon)
\nonumber
\end{array}\right.
\end{equation}
We note, that in the disordered phase (above $T_c(\epsilon)$), no
correction to the density of spin-flips linear in $\epsilon$
occurs.  This is related to the fact that in the weak disorder
limit, the domain size $R$ in Eq.(\ref{energyra}) is limited by
the distance between thermally induced spin-flips, which is too
small to allow for the creation of spin-flips by the random
field.  In this case the lowest-order correction is $\epsilon^2
\beta^4 \exp (\beta I) / 8$.


\subsection{Dominant disorder-induced spin-flips}\label{aRFIM}

%
We now turn to the limit $\exp[-\beta I] \ll \beta^2 \epsilon$,
or equivalently: $z\ll 1$, where the thermal creation of
spin-flips is negligible compared to creation by the random
magnetic field.  To lowest order in $z$, we then have
\cite{Abramowitz} in Eq.(\ref{vaverage}):
\begin{equation}
\frac{z}{\nu}\,\frac{K^{\prime}_{\nu}(z)}{K_{\nu}(z)}\;\simeq\;
-\,\frac{\Gamma(1+\nu)\,\Big(\frac{z}{2}\Big)^{-\nu}\,+\,
\Gamma(1-\nu)\,\Big(\frac{z}{2}\Big)^{\nu}}
{\Gamma(1+\nu)\,\Big(\frac{z}{2}\Big)^{-\nu}\,-\,
\Gamma(1-\nu)\,\Big(\frac{z}{2}\Big)^{\nu}} .
\label{exactsmallz}
\end{equation}
To find the critical temperature, at which $B \rightarrow 0$, it
is sufficient to consider only small values of $\nu\propto B$,
where $\Gamma(1\pm\nu)\simeq e^{\mp\gamma\nu}$ ($\gamma \simeq
0.577$ is Euler's constant), so that Eq.(\ref{exactsmallz})
further reduces to:
\begin{equation}
\frac{z}{\nu}\,\frac{K^{\prime}_{\nu}(z)}{K_{\nu}(z)}\;\simeq\;
-\,\frac{1}{\tanh(\nu\,\ln\frac{2 e^{-\gamma}}{z})} .
\label{gammaapprox}
\end{equation}
Using this result and the definitions Eq.(\ref{besselorder}) and
Eq.(\ref{argument}) of $\nu$ and $z$, respectively, the free energy
Eq.(\ref{vaverage}) finally becomes:
\begin{equation}
\frac{1}{N}\,\left\langle F \right\rangle\;=\;
-\,\frac{\epsilon}{I(T,\epsilon)}\,
\frac{x}{\tanh(x)} .
\label{freesol}
\end{equation}
Here, we introduced
\begin{equation}
I(T,\epsilon)\;=\;
I\,-\,2\,T\,\ln\left(\,\frac{T}{e\,T_{0}}\,\right) ,
\label{ourG}
\end{equation}
with
\begin{equation}
T_0 = T_0(\epsilon) = \sqrt{\,2\,e^{-\gamma-2}\,\epsilon\,} \;,
\label{tnod}
\end{equation}
a characteristic temperature that depends on the strength of the random
magnetic field. Furthermore, we defined $x$,
\begin{equation}
x\;=\;\frac{I(T,\epsilon) B}{\epsilon} \;.
\label{zet}
\end{equation}

At $T = T_0$, the free energy reaches a maximum (cf.  Fig.~1),
implying that the entropy vanishes. The general expression for
the entropy in our model is:
\begin{equation}
\left\langle S \right\rangle \;=\;-\frac{1}{N}\,\frac{\partial
\left\langle F \right\rangle}{\partial T}\;=\;
\frac{2\,\epsilon}{I(T,\epsilon)^{2}}\:
\frac{x^{2}}{\sinh^{2}(x)}
\,\ln\left(\frac{T}{T_{0}}\right) ,
\label{entropy}
\end{equation}
which becomes negative for $T < T_{0}$.  This unphysical behavior
is also observed for isolated chains ($I_{\perp} = 0$)
\cite{Mostovoy97}. Its origin is the continuum approximation,
which breaks down at low temperatures. As we discussed in
Ref.\cite{Mostovoy97}, the problem arises due to the fact that in
the continuum model a spin-flip can take any position on the
chain and is not restricted to the lattice points of the original
discrete model.  The continuum model, therefore, gives incorrect
results as soon as the thermal fluctuation of the spin-flip
positions, $l(T) \sim T^2 / \epsilon$, becomes less than one
lattice constant.  This happens at $T < \sqrt{\epsilon}$, in
agreement with Eq.(\ref{tnod}).  The nature of $T_0$ suggests
that at this temperature the continuum model in fact describes
the zero temperature discrete model.  Using numerical
simulations, we confirmed that for the one-dimensional RFIM this
indeed is a reasonable identification \cite{Mostovoy97}.

Using Eq.(\ref{freesol}), the self-consistency condition
Eq.(\ref{newsc}) now takes the form:
\begin{equation}
\langle\langle\sigma\rangle\rangle =
\frac{2 B}{{\tilde I}_{\perp}} =
\frac{1}{\tanh(x)} - \frac{x}{\sinh^2(x)} .
\label{selfexpl}
\end{equation}
For $\langle\langle\sigma\rangle\rangle \ll 1$, the non-trivial
solution of Eq.(\ref{selfexpl}) for the order parameter reads:
\begin{equation}
\langle\langle\sigma\rangle\rangle\;\propto\;
\sqrt{\,I(T,\epsilon)\,-\,
\frac{3 \epsilon}{{\tilde I}_{\perp}}\,} .
\label{critcurv}
\end{equation}
At the critical temperature $T_{c}(\epsilon)$, the left hand side
of Eq.(\ref{critcurv}) vanishes, yielding
\begin{equation}
\frac{{\tilde I}_{\perp}}{3 \epsilon}\,\left(I\,-\,2\,T_{c}(\epsilon)\,
\ln\Big(\,\frac{T_c}{e T_0(\epsilon)}\,\Big)\right)  = 1 .
\label{tccond}
\end{equation}
Figure~3 shows the critical temperature as a function of the
random magnetic field strength for three values of the interchain
interaction.  The dashed curves are obtained by solving
Eq.(\ref{tccond}).  This yields meaningful results only for
$T_{0}(\epsilon)\leq T_{c}(\epsilon)\leq 0.15 I$, where the upper
limit arises from the fact that the $z\ll 1$ approximation made
in deriving Eq.(\ref{exactsmallz}) starts to break down.  The
solid curves show the results obtained if we do not apply the
small-$z$ expansion, but instead numerically solve the
self-consistency equation Eq.(\ref{newsc}) with the full free
energy Eq.(\ref{vaverage}).  Clearly, over a large disorder
interval, the approximate results give excellent agreement with
the exact ones.  Also presented (dots) are the results in the
weak disorder limit (Eq.(\ref{Tclinear})).  The dash-dotted curve
in Fig.~3 indicates the lower temperature limit
$T_{0}(\epsilon)$, which we argued to correspond to $T=0$ in the
discrete model.

The critical disorder strength $\epsilon_{c}$ below which LRO exists in the
ground state is a function of the interchain interaction and can
be calculated from Eq.(\ref{tccond}) by requiring that
$T_{c}(\epsilon)=T_{0}(\epsilon)$. This yields
\begin{equation}
\epsilon_c \approx \frac{{\tilde I}_{\perp}I}{3}
\left(1+ \frac{c}{9} \sqrt{\frac{3\,{\tilde I}_{\perp}}{I}} \right),
\label{epsilon_c}
\end{equation}
where
\begin{equation}
c = 2 \sqrt{2 e^{-\gamma-2}} \simeq 0.78 .
\label{c}
\end{equation}
It is seen that $\epsilon_c$ is of the order of
${\tilde I}_{\perp}I$.  For a random field strength
$\epsilon\ll \epsilon_{c}$, the ground state ($T=T_{0}$) of the
anisotropic RFIM is nearly perfectly ordered,
$\langle\langle\sigma\rangle\rangle\simeq 1$, as is clear from
Fig.~2.

We now turn to the average density of spin-flips within the
chains, which using Eqs.(\ref{densform}) and (\ref{freesol}), is
obtained as:
\begin{equation}
n_{s}(T,\epsilon) =
\frac{\epsilon}{I^2(T,\epsilon)}\:
\frac{x^2}{\sinh^2(x)} .
\label{densres}
\end{equation}
In the disordered phase ($x = 0$), Eq.(\ref{densres}) reduces to
\begin{equation}
n_s(T,\epsilon) =
\frac{\epsilon}{I^2(T,\epsilon)} .
\label{densresnoint}
\end{equation}
In particular, the latter expression holds for the
one-dimensional RFIM, which is disordered at all temperatures
\cite{Imry75,Imry84}.  Equation (\ref{densresnoint}) was obtained
previously in Ref.~\CITE{Mostovoy97}, where we discussed the
density of kinks in isolated polymer chains.  The temperature
region where Eq.(\ref{densresnoint}) is valid is specified by
Eqs.(\ref{cond1}) and (\ref{cond2}).  However, as we showed in
Ref.~\CITE{Mostovoy97}, the density of spin-flips at zero
temperature in the discrete RFIM is very close to the continuum
result Eq.(\ref{densresnoint}) at $T=T_{0}(\epsilon)$:
\begin{equation}
n_{s}(0,\epsilon) \approx
\frac{\epsilon} {\left(\,I +
c \sqrt{\epsilon} \right)^2} ,
\label{nsdop}
\end{equation}
where $c$ is defined by Eq.(\ref{c}). To lowest
order in $\epsilon$, this result agrees with the estimate
Eq.(\ref{imryma}). The leading correction to Eq.(\ref{imryma}) is
of the order $\epsilon^{3/2}$.

For $\epsilon < \epsilon_c$, LRO does exist and the zero
temperature spin-flip
density should be obtained by numerically solving the order
parameter from Eq.(\ref{selfexpl}) at $T=T_{0}$ and substituting
the result into Eq.(\ref{densres}).  For
${\tilde I}_{\perp}/I=0.02$ the resulting
$\epsilon$-dependence of the spin-flip density is shown in
Fig.~4.  At this particular value of the interchain interaction,
the critical disorder strength is $\epsilon_c \simeq
0.0071I^2$.  It is observed from Fig.~4 that well
below this critical value, the spin-flip density is strongly
suppressed.  In fact, for $\epsilon \ll \epsilon_c$,
Eq.(\ref{densres}) reduces to
\begin{equation}
n_{s}(T_0,\epsilon) \simeq 
\frac{{\tilde I}_{\perp}^{\,2}}{\epsilon}\,
\exp\bigg(-\,
\frac{{\tilde I}_{\perp}\,I(T_{0}, \epsilon)}{\epsilon}\,\bigg) ,
\label{nsop}
\end{equation}
which shows exponential suppression for small $\epsilon$.
Equation (\ref{nsop}) is represented in Fig.~4 by the dashed
curve.  It should be noted that, strictly speaking,
Eq.(\ref{nsop}) is only valid for $\nu = B/(\beta\epsilon) \ll 1$
which was assumed in deriving Eq.(\ref{gammaapprox}) from
Eq.(\ref{exactsmallz}).  However, straightforward analysis shows
that for ${\tilde I}_{\perp}/I=0.02$ and $T=T_0$, this
condition on $\nu$ implies $\epsilon \gg O(10^{-4})$, which holds
for the overwhelming part of Fig.~4.

\section{Discussion}
\label{discussion}

The results for the anisotropic Ising model obtained above admit
a simple interpretation. To this end it is useful to distinguish
between two kinds of spin-flips occuring in chains: kinks and
antikinks. For kinks, the spins are positive to the left of the
spin-flip and negative to the right. For antikinks, the opposite
holds. Obviously, kinks and antikinks have the same creation
energy, and along the chain always an alternation of kinks and
antikinks occur.

We first discuss the density of thermally induced spin-flips in
the absence of a random magnetic field (Eq.(\ref{denssimpl}) with
$\epsilon = 0$).  As we assumed the temperature to be much
smaller than the spin-flip creation energy, $I$, the density of
the thermally-induced spin-flips is small.  In the disordered
phase (above $T_c$), one can neglect the correlations between the
positions of spin-flips in different chains and the density of
the spin-flips (kinks and antikinks) is given by the Boltzmann
formula: $n_s = e^{-\beta I}$.  In the ordered phase, the
interchain interactions bind kinks and antikinks within one
chain into pairs, which is clear from the fact that 
$n_s \propto e^{- 2 \beta I}$ below $T_c$.  As follows from 
Eq.(\ref{action}) the binding energy $V$ of such a pair is 
proportional to the distance $R$ between kink and antikink:
\begin{equation}
\label{string}
V(R) = {\tilde I}_{\perp} R ,
\label{V(R)}
\end{equation}
as over this distance the spin sign is opposite to
$\langle\langle\sigma\rangle\rangle$. Thus, below $T_c(0)$, the
density of the thermally induced kink-antikink pairs is:
\begin{equation}
n_{pair} \sim {\bar R} (T) e^{- 2 \beta I}\;,
\end{equation}
where the factor ${\bar R} (T) = T / ({\tilde I}_{\perp})$ - the
average size of the pair at temperature $T$ - comes from the
sum over all possible distances between kinks and antikinks in
the partition function.  The density of spin-flips in the ordered
phase equals $2 n_{pair}$, in accordance with Eq.(\ref{denssimpl}).

The transition between the ordered and disordered phases occurs
when the kink-antikink pairs dissociate into free kinks and
antikinks. Thus, at the critical temperature the average length of the
kink-antikink pair is of the order of the average distance $1 /
n_s$ between spin-flips,
\begin{equation}
{\bar R}(T_c(0)) \sim \exp\left(\frac{I}{T_c(0)}\right) ,
\end{equation}
which explains the physical content of Eq.(\ref{tcsimpl})
\cite{Khomskii96}.

Next we discuss the density of spin-flips in the disordered phase
induced predominantly by the random magnetic field  [see
Eq.(\ref{densresnoint})].  This equation coincides with the
simple Imry and Ma estimate Eq.(\ref{imryma}), except that the
kink creation energy, $I$, is replaced by $I(T, \epsilon)$
defined in Eq.(\ref{ourG}).  This can be understood as follows:
The spin-flips enter into the partition function with the weight:
$w(T) = l(T) e^{ - \beta I}$, where $l(T) \sim T^2 / \epsilon$ is
the thermal fluctuation of the spin-flip position in the presence
of the random magnetic field. We can now write the weight in the
form: $w(T) = e^{ - \beta I(T,\epsilon)}$, where $I(T,\epsilon) =
I - T \ln l(T)$ is the spin-flip free energy, which also includes
the entropy of the spin-flip location. It is clear that the free
energy, rather than the ``bare'' kink creation energy, $I$,
should enter into the correct expression Eq.(\ref{densresnoint})
for the density of spin-flips.

Finally, we discuss the critical strength of the random magnetic
field, given by Eq.(\ref{epsilon_c}), above which the system is
disordered at all temperatures.  In an isolated chain, spin-flips
are induced by an arbitrarily weak random magnetic field [see
Eq.(\ref{densresnoint})], because the distances between the
spin-flips can be made as large as is necessary to compensate for
the spin-flip creation energy by the energy of the interaction
with the random magnetic field.  In the presence of interchain
interactions, however, the distances between spin-flips are not
allowed to be arbitrarily large, because of the potential
Eq.(\ref{V(R)}).  The potential grows linearly with the distance
$R$ between kink and antikink, while the energy gain due to
kink-antikink pair creation grows proportional to $\sqrt{R}$.
Thus, in the presence of interchain interactions, the equation
describing the balance between the energy gain and energy loss
due to the creation of a kink-antikink pair of size $R$ [cf.
Eq.(\ref{energyra})],
\begin{equation}
2 I + {\tilde I}_{\perp} R \sim 2 \sqrt{\epsilon R}\;,
\label{balance}
\end{equation}
only has a solution if $\epsilon > \epsilon_c \sim {\tilde
I}_{\perp} I$, in accordance with Eq.(\ref{epsilon_c}).  For
$\epsilon < \epsilon_c$ the distances between disorder-induced
spin-flips in isolated chains are very large, so that they are
suppressed by the interchain interactions and the system at zero
temperature is in the ordered state.  If, on the other hand,
$\epsilon > \epsilon_c$ the system is disordered at all
temperatures.

The dependence of the critical temperature on the disorder
strength in the RFIM can be compared with that in another
Ising-type model of disordered (spin-) Peierls systems considered
in Ref.~\cite{Khomskii96}.  There it was assumed that impurities
randomly cut the chains into finite segments, some of which, {\em
e.g.}, the ones with an odd number of units, contain at least one
soliton.  This was modeled by an ensemble of finite Ising
chains whose length are taken from a Poisson distribution 
and which are subjected to a mean field and four different kinds 
of boundary conditions.  For small impurity concentration $x$, the 
critical temperature in that model decreases linearly with $x$, 
just as in the RFIM [see Eq.(\ref{Tclinear})].  On the other hand, 
for large $x$, $T_c(x)\propto 1/x$ and at zero temperature the 
ground state is ordered for all values of $x$.  The absence of 
a critical concentration, above which LRO is destroyed, is related 
to the fact that in that model the positions of spin-flips
(solitons) within the finite chains are not fixed.  At $T=0$ the
spin-flips are located near one of the chain ends to minimize the 
size of ``wrong'' domains.  In the RFIM, however, the distances 
between neighboring spin-flips are governed by the energy balance 
Eq.(\ref{balance}) and cannot be made arbitrarily small, which 
results in the existence of a critical random field strength.


\section{Summary and Conclusions}\label{concl}

In this paper, we studied the anisotropic RFIM, consisting of
spin chains with interchain exchange interactions that are much
weaker than the intrachain interactions.  Treating the interchain
interactions in the mean-field approximation, we found an
analytical solution Eq.(\ref{vaverage}) for the free energy.  The
crux of our approach lies in the fact that in this chain
mean-field approximation the free energy of the continuum version
of the model can be related to a one-dimensional Brownian motion.
We found the stationary solution of the corresponding
Fokker-Planck equation, which enabled us to perform the average
of the free energy over the random magnetic field realizations.
From this, we obtained the order parameter as a function of
temperature and random field strength and, more specifically, the
critical temperature below which LRO occurs as a function of the
random field strength.  We also calculated the density of
spin-flips in the chains.  As we decribed in
Sec.~\ref{discussion}, our results have a clear physical
interpretation.

The chain mean-field approximation restricts the validity of our
results to $d > 2$, where thermodynamic quantities suffer much
less from fluctuations than in lower dimensional systems.  We
believe that the critical temperature $T_{c}(\epsilon)$ given by
Eq.(\ref{tccond}) describes the phase transition in three
dimensions qualitatively correct and is quantitatively
correct for a strongly anisotropic realization of the RFIM.
Clearly, our mean-field treatment of interchain interactions is
not applicable in two dimensions, where LRO is lacking for all
temperatures and (or) random-field strengths \cite{Imbrie84}.

The one-dimensional RFIM had already been solved by us in the
context of disorder-induced solitons in a single chain of the
polymer trans-polyacetylene \cite{Mostovoy97}.  Our present
treatment of the anisotropic $d$-dimensional RFIM allows us to
extend that study to include the soliton-antisoliton confinement
that is imposed by electron hopping between neighboring
polyacetylene chains in the bulk material.  In particular, the
exponential suppression of the density of spin-flips well below
the critical random field strength (see Eq.(\ref{nsop})) may be
relevant to explain the absence of experimental signatures of
these solitons.  In a forthcoming paper \cite{Figge97}, we will
focus more particularly on this application of the anisotropic
RFIM and discuss various experimental observables.
We finally note that if $B$ is considered an external (rather
than self-consistent) field, our model describes the lattice
dimerization of
a disordered polymer chain with a non-degenerate ground state
\cite{SSHK}.
$B$ then is proportional to the energy difference
between the two phases with different signs of the dimerization.

\section*{Acknowledgment}
This work is part of the research program of the Stichting
Fundamenteel Onderzoek der Materie (FOM), which is financially
supported by the Nederlandse Organisatie voor Wetenschappelijk
Onderzoek (NWO).

%

%

%
%
%
\newpage
\begin{center}
{\bf Figure captions}
\end{center}
\vspace{1cm}

Fig.~1. The average free energy of the anisotropic RFIM as a
function of temperature for three values of the mean field:
$B = 0.005I$, $0.01I$, and  $0.02I$ (curves a, b, and c, 
respectively). Solid curves represent the analytical result 
Eq.(\ref{vaverage}) for the continuum approximation to the 
RFIM; dashed curves are obtained from numerical simulations 
of the RFIM. In all cases, the random field strength was taken 
$\epsilon = 0.01I^2$. The thin line at $T = T_0$ gives the 
position of the maxima of the analytical solution.

\bigskip

Fig.~2.  The order parameter
$\langle\langle\sigma\rangle\rangle$ of the ansitropic RFIM
as a function of the random magnetic field strength $\epsilon$
and temperature $T$ (for $T > T_{0}(\epsilon)$). The plot was 
obtained by solving the self-consistency condition 
Eq.(\ref{newsc}) with the analytical solution Eq.(\ref{vaverage}) 
for the free energy. The interchain interaction was chosen such 
that ${\tilde I}_{\perp}/I$ = 0.02, in which case $T_{c} \simeq
0.295I$ (at $\epsilon = 0$) and $\epsilon_{c} = 0.0071 I^{2}$. 
At $(T,\epsilon) = (0,0)$, the order parameter reaches its
maximum: $\langle\langle\sigma\rangle\rangle = 1$.
The thick curve in the $\langle\langle\sigma\rangle\rangle = 0$ 
plane separates the ($\epsilon,T$) regions with and without LRO.

\bigskip

Fig.~3.  The critical temperature $T_{c}(\epsilon)$ of the
anisotropic RFIM as a function of the random magnetic field
strength for ${\tilde I}_{\perp}/I$ = 0.002, 0.01, and 0.02 
(curves a, b, and c, respectively). Solid curves derive from
solving Eq.(\ref{newsc}) with the exact free energy 
Eq.(\ref{vaverage}). The dashed curves are obtained by solving
Eq.(\ref{tccond}), while the dots represent the small-$\epsilon$
behaviour Eq.(\ref{Tclinear}). The dash-dotted curve shows the 
temperature $T_{0}(\epsilon)$, below which the continuum 
approximation breaks down and which corresponds approximately to 
$T=0$ in the discrete RFIM.

\bigskip

Fig.~4.  Density of spin-flips within the chains of the
ansitropic RFIM as a function of the random magnetic field
strength at $T=T_{0}$, {\em i.e.}, in the ground state.  The
interchain interaction was chosen such that
${\tilde I}_{\perp}/I$ = 0.02.  The dashed line shows the
analytical result Eq.(\ref{nsop}) for
$\epsilon\ll\epsilon_{c}\simeq 0.0071I^2$.
\end{document}